\documentclass[aps,prd,twocolumn,eqsecnum,showpacs,amsmath]{revtex4}

\usepackage{subfigure}
\usepackage[dvips]{color,graphicx}
\usepackage{amsfonts,amssymb,theorem}
\newcommand{\D}{{\rm d}}
\newcommand{\vect}[1]{\!\!\!\mbox{ \boldmath $#1$}}

{\theorembodyfont{\upshape}
}
{\theorembodyfont{\upshape}
}
{\theorembodyfont{\upshape}
}
{\theorembodyfont{\upshape}
}
{\theorembodyfont{\upshape}
}
{\theorembodyfont{\upshape}
}

\newcommand{\dalm}{\kern1pt\vbox{\hrule height 0.9pt\hbox{\vrule width
0.9pt\hskip 2.5pt\vbox{\vskip 5.5pt}\hskip 3pt\vrule width 0.3pt}\hrule height
0.3pt}\kern1pt}

\def\b2hat{ {\hat b}_2 }

\begin{document}
%\thispagestyle{empty}

%<<<<<<<<<<<<< TITLE >>>>>>>>>>>>>>>%
\title{
Creation of the universe with a stealth scalar field
}

%<<<<<<<<<<<<< AUTHOR >>>>>>>>>>>>>>>%
\author{Hideki Maeda${}^{a}$}
\email{hideki-at-cecs.cl}
\author{Kei-ichi Maeda${}^{b,c}$}
\email{maeda-at-waseda.jp}

%<<<<<<<<<<<<< ADDRESS >>>>>>>>>>>>>>>%

\address{ 
	${}^a$ Centro de Estudios Cient\'{\i}ficos (CECs), Casilla 1469,
 Valdivia, Chile \\
	${}^b$ Department of Physics, Waseda University, Okubo 3-4-1, 
Shinjuku, Tokyo 169-8555, Japan.\\
	${}^c$ DAMTP, University of Cambridge, Wilberforce Road, 
Cambridge CB3 0WA, U.K.
}

%<<<<<<<<<<<<< DATE >>>>>>>>>>>>>>>%
\date{\today}

%======================================%
%<<<<<<<<<<<<< ABSTRACT >>>>>>>>>>>>>>>%
%======================================%
\begin{abstract} 
The stealth scalar field is a non-trivial configuration without any back-reaction to geometry, which is characteristic for non-minimally coupled scalar fields.
Studying the creation probability of the de Sitter universe with a stealth scalar field by the Hartle and Hawking's semi-classical method, we show that the effect of the stealth field can be significant.
For the class of scalar fields we consider, creation with a stealth field is possible for a discrete value of the coupling constant and its creation probability is always less than that with a trivial scalar field.
However, those creation rates can be almost the same depending on the parameters of the theory.
\end{abstract}

%<<<<<<<<<<<<< PACS NUMBER >>>>>>>>>>>>>>>%
\pacs{
04.60.Bc 	%Phenomenology of quantum gravity 
98.80.Qc 	%Quantum cosmology
} 
% CECS-PHY-12/06
\maketitle

%======================================%
%<<<<<<<<<<<< SECTION I  >>>>>>>>>>>>>>%
%======================================%
\section{Introduction}
\label{sec1}
At present, the framework of the Big-Bang cosmology with an early inflationary
 era achieves a clear consensus from cosmologists.
Although the number of the first observational data by Hubble advocating the
 expanding universe was small~\cite{hubble1929}, a variety of current 
observations supports the assertion.
On the other hand, another fact that the universe is spatially homogeneous and
 isotropic in a large scale indicates that 
the Friedmann-Lema{\^ i}tre-Robertson-Walker (FLRW) cosmological spacetime may
 be a good model as the zeroth-order approximation of our universe.
A direct consequence of these two is that there was an extremely dense and hot
 era in the history of our universe, called the Big Bang.

While the great success of the Big-Bang cosmology is 
the prediction of the 
observed cosmic microwave background (CMB) and the production of the
 light elements, several problems are
 left unsolved in this framework.
Among them, the flatness problem, horizon problem, and monopole problem
(if one believes grand unified theories of fundamental interactions) are
 resolved by an additional but simple assumption; there was a phase with 
accelerating expansion before the Big-Bang era.
This assumption of the cosmic inflation suits the observational data of 
the CMB quite well and completes the modern 
framework of the Big-Bang cosmology.

Even in this framework, however, the initial singularity problem remains
 unsolved and clearly shows the limit of the classical description of the 
early universe.
The Big-Bang singularity is a consequence of the singularity theorems but the
 strong energy condition is assumed to prove them in general relativity~\cite{he}.
By this reason, one may think that inflation is able to slip through the net
 of the singularity theorems.
However, under certain reasonable conditions without energy conditions, 
it was shown that the inflationary universe must have a past spacetime
 boundary, namely the initial moment of the classical universe~\cite{bgv2003}.

It is reasonable to assume that the quantum aspect of gravity dominates close 
to the initial singularity.
This leads us to quantum cosmology, which is a
 branch of the minisuperspace approach of canonical quantum 
gravity~\cite{bv2010}.
In quantum cosmology, the Wheeler-de~Witt equation, obtained by the 
quantization of the Hamiltonian constraint of Einstein equations, determines
 the wave function of the universe.
In this context, Vilenkin
proposed the spontaneous creation of the universe by 
quantum tunneling from nothing~\cite{vilenkin1982,vilenkin}.
This picture is understood as the creation of the universe in the de~Sitter 
space, represented by the Euclidean solution called instanton.
Because the mathematical description is analogous to the quantum tunneling 
through the potential barrier, it is claimed that the creation probability is 
proportional to  $e^{-|S_{\rm E}|/\hbar}$~\cite{linde},
 where $S_{\rm E}$ is 
the value of the Euclidean action evaluated for the instanton solution.
On the other hand, Hartle and Hawking made the no-boundary proposal claiming
 that the wave function of the universe is obtained by a path integral over 
non-singular compact Euclidean spaces~\cite{hh1983,hhh}, by which 
they evaluated the creation probability as $e^{-S_{\rm E}/\hbar}$.

The simplest vacuum instanton solution is the Hawking-Moss de~Sitter instanton 
which is nothing but an $S^4$ in the 4D Euclidean space
 and can be analytically continued to 
the Lorentzian de~Sitter universe with the spatially closed 
slicing~\cite{hm1982}.
While the value of the Euclidean action in vacuum is simply proportional 
to the volume of the instanton space, the contribution of matter to the 
creation probability is non-trivial.
Non-minimally coupled scalar fields are interesting in this context because 
they allow a ``stealth'' configuration, namely a no-trivial solution which
 does not give any back-reaction to gravity.
Such a configurations was already recognized in 1976 at the latest both 
for a conformally coupled scalar field and a Yang-Mills field in the Minkowski 
spacetime~\cite{dAFF1976}.
More recently, the stealth solution was found in the three-dimensional 
(locally) anti-de~Sitter
 (AdS) spacetime~\cite{agmp2000} and also in the Minkowski~\cite{AyonBeato:2005tu}, de~Sitter (dS)~\cite{bjj2008}, and AdS spacetimes~\cite{Martinez:2005di} in four dimensions.
(See also~\cite{stealth,deHaro:2006nv}.)
Then a natural question arises: if a de~Sitter instanton is possible both 
with a trivial scalar field and with a stealth scalar field, which universe 
is preferred at the moment of creation?
In this paper, we consider this problem.

In the following section, we present our system and obtain a stealth solution 
in the de~Sitter instanton space.
In Section~III, we study the creation probability of the de~Sitter universe
 based on our solutions.
Concluding remarks and discussions are summarized in Section~IV.
Our basic notation follows \cite{wald}.
The conventions for curvature tensors are
$[\nabla _\rho ,\nabla_\sigma]V^\mu ={R^\mu }_{\nu\rho\sigma}V^\nu$
and $R_{\mu \nu }={R^\rho }_{\mu \rho \nu }$.
The signature of the Minkowski spacetime is $(-,+,+,+)$ and
Greek indices run over all spacetime indices.
We adopt the units such that $c=\hbar=8\pi G=1$.

%======================================%
%<<<<<<<<<<<< SECTION I  >>>>>>>>>>>>>>%
%======================================%
\section{Stealth scalar field in de~Sitter space}
\subsection{System}
We consider general relativity coupled with a scalar field non-minimally, 
of which action is given by 
\begin{align}
\label{action}
S=&S_{\rm g}+S_{\phi},
\end{align}
where 
\begin{align}
S_{\rm g}=&\frac{1}{2}\int d^4x\sqrt{-g}(R-2\Lambda),\\
S_{\phi}=&-\int \D^4x \sqrt{-{ g}} \left[\frac12({\nabla} \phi)^2 
+\frac12\xi R\phi^2+V(\phi)  \right].\label{action-con}
\end{align}
$S_{\phi}$ is the action for a non-minimally coupled scalar field.
The resulting field equations are
\begin{align}
&G_{\mu\nu}+\Lambda g_{\mu\nu}= T^{(\phi)}_{\mu\nu},\\
&\nabla^2\phi -\xi R\phi -\frac{dV}{d\phi}=0,\label{KG}
\end{align}
where  
\begin{align}
T^{(\phi)}_{\mu\nu} =&(1-2\xi)(\nabla _\mu \phi)(\nabla_\nu\phi )
+\biggl(2\xi -\frac12\biggl)(\nabla \phi)^2 g_{\mu\nu}
\nonumber \\
&-Vg_{\mu\nu} +\xi \phi ^2 G_{\mu\nu}+2\xi \phi (-\nabla_\mu\nabla_\nu\phi +g_{\mu\nu}
\nabla^2\phi ).
\end{align}
For a conformally coupled scalar field, the coupling constant $\xi$ and 
the potential are chosen as $\xi=1/6$ and $V(\phi)=\alpha \phi^{4}$, where 
$\alpha$ is a constant, and then the trace of $T^{(\phi)}_{\mu\nu}$ 
is vanishing.

\subsection{Stealth scalar-field solution}
Here we consider the Euclidean de Sitter space:
\begin{align}
ds^2=&d\tau^2+a(\tau)^2\biggl(d\chi^2+\sin^2\chi(d\theta^2+\sin^2\theta 
d\varphi^2)\biggl),\label{EdS} \\
a(\tau)=&\frac{1}{H}\cos H\tau,
\end{align}
where $H:=\sqrt{\Lambda/3}$ and the domain of $\tau$ is $-\pi/(2H)\le \tau\le \pi/(2H)$.
The domains of other coordinates are $0 \le \chi\le \pi$, $0 \le \theta\le \pi$, and $0 \le \varphi\le 2\pi$.
We obtain a stealth scalar field in this spacetime, namely a non-trivial 
scalar field with $T^{(\phi)}_{\mu\nu}\equiv 0$.
(The AdS instanton with a stealth conformally coupled scalar field was 
obtained in~\cite{deHaro:2006nv}.)
The field equations for the stealth scalar field are
\begin{align}
0 =&(1-2\xi)(\nabla _\mu \phi)(\nabla_\nu\phi )
+\biggl(2\xi -\frac12\biggl)(\nabla \phi)^2 g_{\mu\nu}-Vg_{\mu\nu} 
\nonumber \\
&-\xi \Lambda \phi ^2g_{\mu\nu}+2\xi \phi (-\nabla_\mu\nabla_\nu\phi 
+g_{\mu\nu}\nabla^2\phi ),\\
0=&\nabla^2\phi -4\xi \Lambda\phi -\frac{dV}{d\phi}
\,,
\end{align}
where we have used the Einstein equations with vanishing energy-momentum 
tensor: $G_{\mu\nu}+\Lambda g_{\mu\nu}=0$ and $R=4\Lambda$.

Assuming $\phi=\phi(\tau)$, 
we find basic equations:
\begin{align}
&
{1\over 2}(\phi')^2-V-\xi \Lambda\phi^2-6\xi H \tan H\tau \phi \phi'=0,
\label{00_compo}\\
&(4\xi-1)(\phi')^2-2V-2\xi \Lambda\phi^2 \nonumber \\
&~~~~~~~~~~~~+4\xi\phi\left(
\phi''-2H\tan H\tau \phi'\right)=0,
\label{ii_compo}\\
&
\phi''-3H\tan H\tau \phi'-4\xi \Lambda\phi-\frac{dV}{d\phi}=0
\label{scalar_eq}
\,.
\end{align}
From Eqs. (\ref{00_compo}) and (\ref{ii_compo}), we find
\begin{align}
(2\xi-1)(\phi')^2+2\xi\phi\left(
\phi''+H\tan H\tau \phi'\right)=0
\,,
\end{align}
from which,
we obtain a general solution for the case of $\xi\ne 1/4$:
\begin{align}
\phi=&(p\sin H\tau +q)^{n},\label{solution}
\end{align}
where $p$ and $q$ are constants and $n$ is defined by 
\begin{align}
n:=\frac{2\xi}{4\xi-1}\quad \longleftrightarrow \quad \xi=\frac{n}{2(2n-1)},\label{xi-const}
\end{align} 
Inserting the solution (\ref{solution}) into (\ref{00_compo}), 
we find the corresponding potential $V$ as
\begin{align}
\label{potential-3}
V(\phi)=&\Lambda\phi^2\biggl(A\phi^{(1-4\xi)/\xi}+B\phi^{(1-4\xi)/2\xi}+C\biggl)  \\
=&\Lambda\phi^2\biggl(A\phi^{-2/n}+B\phi^{-1/n}+C\biggl),\label{potential-3-2}
\end{align} 
where 
\begin{align}
A&:=\frac{2\xi^2(p^2-q^2)}{3(4\xi-1)^2}=\frac{n^2}{6}(p^2-q^2),\label{A-2}
\\
B&:=\frac{8q\xi^2(6\xi-1)}{3(4\xi-1)^2}=\frac{2n^2(n+1)q}{3(2n-1)}, \label{B-2} 
\\
C&:=-\frac{\xi(6\xi-1)(16\xi-3)}{3(4\xi-1)^2}=-\frac{n(n+1)(2n+3)}{6(2n-1)}\,.\label{C-2}
\end{align} 
The coupling constant $\xi=1/6$ for a conformally coupled scalar field corresponds to $n=-1$ and then the potential becomes very simple:
\begin{align}
V(\phi)=\Lambda A\phi^4.
\end{align}

The constants in the solution (\ref{solution}) are given by 
\begin{align}
p^2&=\frac{3(4\xi-1)^2}{2\xi^2}\biggl[A+\frac{3(4\xi-1)^2B^2}{32\xi^2(6\xi-1)^2}\biggl] \\
&=\frac{6}{n^2}\biggl[A+\frac{3(2n-1)^2B^2}{8n^2(n+1)^2}\biggl],\\
q&=\frac{3(4\xi-1)^2 B}{8\xi^2(6\xi-1)}=\frac{3(2n-1)B}{2n^2(n+1)}\,.
\end{align} 
The coupling constant $\xi$ in the non-minimal coupling must be related to the constant $C$ as (\ref{C-2}).
It is noted that the corresponding potential is quadratic for the de~Sitter
 spacetime with the spatially flat slicing~\cite{bjj2008}.

In the case of $\xi=1/4$, the stealth solution is given by 
\begin{align}
\phi=&K e^{w\sin H\tau},
\label{sol_1/4}
\end{align}
where $K$ and $w$ are constants.
The corresponding potential is given by 
\begin{align}
V(\phi)=\Lambda\phi^2\biggl[{\bar A}+{\bar B}\ln {\phi}-\frac{1}{6}(\ln{\phi})^2\biggl]\,.
\end{align} 
The constants in the solution are given by 
\begin{align}
K=
e^{3({\bar B}+1/2)}, \quad w^2=3\biggl(2{\bar A}-{\bar B}-\frac{1}{2}\biggl).\label{paras}
\end{align}
We assume the values of ${\bar A}$ and ${\bar B}$ such that $w^2>0$ holds.

\subsection{Instanton}
\label{sec:instanton}
Here we shall impose regularity conditions for the scalar field in order to find a regular instanton solution on de Sitter space.
We require (I) finiteness of $\phi$ and (II) $\phi'=0$ at the de Sitter poles $\tau=\pm \pi/(2H)$.

In the case of $\xi\neq 1/4$, the following expression
\begin{align}
\phi'=&n pH\cos H\tau(p\sin H\tau +q)^{n-1} \label{dsolution}
\end{align}
shows that the regularity conditions are satisfied in three cases; (i) $q>|p|$, (ii) $q=|p|$ with $n>1/2$, and (iii) $q< |p|$ and $n$ is a positive integer.
In the cases (i) and (ii), $\phi$ is non-negative.
In the case (iii), $\phi$ vanishes at some latitude and changes the sign there if $n$ is odd, while the sign is fixed for even $n$.
Because $\phi$ must not be negative if the power exponent $2/n$ in the potential (\ref{potential-3-2}) is non-integer, $n=1$ or even $n$ is required in the case (iii).

In the case of $\xi=1/4$, the scalar field 
(\ref{sol_1/4}) is regular at the coordinate boundary $\tau =\pm\pi/(2H)$.
So it gives an instanton solution.

\subsection{de Sitter solution with a constant scalar field}
\label{sec:dS}
Now let us discuss whether the present system also allows a de~Sitter solution with a trivial scalar field, namely $\phi =\phi_{\rm c}=$constant.
Equation~(\ref{KG}) gives an algebraic equation for $\phi_{\rm c}$:
\begin{align}
0=& \phi_{\rm c}\biggl(\frac{n-1}{n}A\phi_{\rm c}^{-2/n}+\frac{2n-1}{2n}B\phi_{\rm c}^{-1/n}+C+\frac{n}{2n-1} \biggl) \label{alg1}
%=& \frac{n}{3}\phi_{\rm c}\biggl(\frac{n-1}{2}(p^2-q^2)\phi_{\rm c}^{-2/n}+(n+1)q\phi_{\rm c}^{-1/n}-\frac{n+3}{2}\biggl) 
\end{align}
for $\xi\ne 1/4$ and 
\begin{align}
0=&\phi_{\rm c}\biggl(3+6{\bar A}+3{\bar B}+(6{\bar B}-1)\ln {\phi}_{\rm c}-(\ln{\phi}_{\rm c})^2\biggl) \label{alg1-2}
\end{align}
for $\xi=1/4$.
The corresponding energy-momentum tensor is given by 
\begin{align}
T^{(\phi)}_{\mu\nu} =&-\Lambda\phi_{\rm c}^2\biggl(A\phi_{\rm c}^{-2/n}+B\phi_{\rm c}^{-1/n}+C+\frac{n}{2(2n-1)}\biggl)g_{\mu\nu} \label{alg2}
\end{align}
for $\xi\ne 1/4$ and 
\begin{align}
T^{(\phi)}_{\mu\nu} =&-\Lambda\phi_{\rm c}^2\biggl(\frac{1}{4}+{\bar A}+{\bar B}\ln {\phi}_{\rm c}-\frac16(\ln{\phi}_{\rm c})^2\biggl)g_{\mu\nu} \label{alg2-2}
\end{align}
for $\xi=1/4$.
Here we focus on the solution with $T^{(\phi)}_{\mu\nu} \equiv 0$ and $p,n\ne 0$ in order to compare with the stealth solution in the subsequent section. 

In the case of $\xi \ne 1/4$, Eqs.~(\ref{alg1}) and (\ref{alg2}) admit a solution $\phi_{\rm c}=0$ in two cases; (i) $q\ne |p|$ with $n>2$ and (ii) $q=|p|$ with $n>1$. 
On the other hand, inside the large brackets of Eqs.~(\ref{alg1}) and (\ref{alg2}) are quadratic equations for $x:=\phi_{\rm c}^{-1/n}$, which admit a non-zero solution $\phi_{\rm c}=(q^2-p^2)^{-1/2}$ only if $n=-1$ and $|q|>|p|$.

In the case of $\xi =1/4$, Eqs.~(\ref{alg1-2}) and (\ref{alg2-2}) admit a solution $\phi_{\rm c}=0$.
They also admit a non-zero solution $\ln \phi_{\rm c}=3{\bar B}+3/2$ only if ${\bar A}=-3{\bar B}^2/2+1/8$, but it is discarded because Eq.~(\ref{paras}) gives a contradiction $w^2=-(3{\bar B}+1/2)^2-1/2<0$.

%======================================%
%<<<<<<<<<<<< SECTION I  >>>>>>>>>>>>>>%
%======================================%
\section{Creation of the universe}
When we discuss creation of the universe by use of an instanton
solution, the constraint on the solution becomes much more tight.
We will see that the theory with certain values of parameters admits creation of the universe with and without a stealth scalar field.
In this section, we discuss which configuration is preferred at the moment of the universe creation by the use of the Hartle and Hawking's semi-classical instanton approach.

\subsection{Closed universe}
First we consider creation of the de~Sitter universe in the spatially closed slicing.
The de Sitter universe can be quantum mechanically created 
at $\tau=0$, where $a'$ vanishes and then the spacetime can be 
analytically continued by the Wick rotation $\tau=iT$.
Since the scalar field is also associated to the de~Sitter spacetime,
$\phi'$ also vanishes at $\tau=0$.
For $\xi \ne 1/4$, it is possible only in the case of $q=0$.
The case of  $\xi=1/4$ is not possible to discuss the creation of the universe because $\phi'$ does not vanish at  $\tau=0$.

The scalar field and the corresponding potential for $\xi \ne 1/4$ are now given by 
\begin{align}
\phi=&\phi_0(\sin H\tau)^{n},\label{stealth}\\
V(\phi)=&\Lambda \phi^2\biggl(A\phi^{-{2\over n}}+C\biggl)
\label{pot1}
\,,
\end{align} 
where 
\begin{align}
\phi_0:=&p^{n},\\
A=& \frac{p^2n^2}{6},
\label{pot1_A}
\\
C=& -\frac{n(n+1)(2n+3)}{6(2n-1)}
\label{pot1_C}
\,.
\end{align} 
The scalar field for $\xi \ne 1/4$ in the de~Sitter region is given by 
\begin{align}
\phi=(ip\sinh HT)^{n}.
\end{align}
In order for the scalar field to be real, $n$ must be a positive even integer.
In addition, because $\phi$ must not be negative if the power exponent in the potential $2/n$ is non-integer, the creation of the universe is allowed only for $n=2$ or $n=4k~(k\in \mathbb{N})$, which also satisfies the regularity conditions considered in Sec.~\ref{sec:instanton}.

The domain of $\phi$ for the instanton is given by $0\le \phi\le \phi_0$.
The first term in the potential dominates in the region of small $\phi$
 where the potential is positive.
Because of
\begin{align}
V(\phi_0)=&-\frac{\Lambda n(2n+1) \phi_0^{2}}{2(2n-1)}<0,
\end{align} 
the potential has a negative domain.

We will compare the creation probabilities of two de~Sitter universe with the same value of $\Lambda$; one is with a stealth scalar field and the other is with a constant scalar field.
We will consider only the case of $n=4k~(k\in \mathbb{N})$ because the result in Sec.~\ref{sec:dS} shows that, for $n=2$, the theory does not admit a de~Sitter solution with the same value of $\Lambda$ and a constant scalar field.

\subsection{Open universe}
The Euclidean de Sitter space (\ref{EdS}) can be also continued to the de~Sitter universe in the spatially open slicing.
To see this, we first shift the origin of th Euclidean time as $\tau={\bar \tau}-\pi/(2H)$, with which the Euclidean de Sitter space (\ref{EdS}) is written as
\begin{align}
ds^2=&d{\bar \tau}^2+b({\bar \tau})^2(d\chi^2+\sin^2\chi d\Omega_2^2),\label{EdS2}
\end{align}
where $b({\bar \tau})=(1/H)\sin H{\bar \tau}$ and the domain of ${\bar \tau}$ and $\chi$ are $0\le {\bar \tau}\le \pi/H$ and $0 \le \chi\le \pi$, respectively.
The stealth solution is given by 
\begin{align}
\label{ste-3}
\phi= \left\{
\begin{array}{ll}
(-p\cos H{\bar \tau}+q)^{n} & (\xi\ne 1/4),\\
K e^{-w\cos H{\bar \tau}} & (\xi= 1/4).
\end{array} \right. 
\end{align}

One obtains an open universe by continuing $\chi$ such that $\chi$ runs from $0$ to $\pi/2$ in the Euclidean region and then in the imaginary direction in the Lorentzian region.
Defining a time coordinate by $\chi=\pi/2+i{\bar t}$, we obtain the de~Sitter metric in the following form:
\begin{align}
ds^2=&d{\bar \tau}^2+b({\bar \tau})^2(-d{\bar t}^2+\cosh^2{\bar t}d\Omega_2^2),\label{EdS3} 
\end{align}
where the domain of ${\bar t}$ in this region is $0<{\bar t}<\infty$ and the scalar field (\ref{ste-3}) is constant in time.
Since the universe (\ref{EdS3}) is created at $\chi=\pi/2$ and hence the boundary condition for $\phi$ is $d\phi/d\chi(\chi=\pi/2)=0$, 
which is trivially satisfied. 
The scale factor $b({\bar \tau})$ vanishes at ${\bar \tau}=0$ and ${\bar \tau}=\pi/H$.
They are coordinate singularities and the spacetime is analytic there.
We emphasize that this spacetime is not the same as the one obtained from the Hawking-Turok instanton~\cite{ht1998}, which contains a singularity.

The metric (\ref{EdS3}) is singular at the null surface ${\bar \tau}=0$ (and also ${\bar \tau}=\pi/H$) but it is a coordinate singularity and can be analytically extended beyond there.
As shown below, the de~Sitter metric in the extended region is written in the spatially open slicing.
By the coordinate transformations
\begin{eqnarray}
u=\frac{1}{\sqrt{2}}{\bar \tau}e^{-{\bar t}}, ~~~v=-\frac{1}{\sqrt{2}}{\bar \tau}e^{{\bar t}}, \label{trans-1a}
\end{eqnarray}
of which inverse is
\begin{eqnarray}
{\bar t}={\rm arctanh}\biggl(\frac{v+u}{v-u}\biggl), ~~~{\bar \tau}=\sqrt{-2uv}, \label{trans-1b}
\end{eqnarray}
the metric (\ref{EdS3}) is transformed into 
\begin{align}
ds^2=&-\frac{(vdu+udv)^2}{2uv}-\frac{\sin^2(H\sqrt{-2uv})}{H^2}\frac{(vdu-udv)^2}{4u^2v^2} \nonumber \\
&-\frac{\sin^2(H\sqrt{-2uv})}{H^2}\frac{(v-u)^2}{4uv}d\Omega_2^2.\label{EdS4} 
\end{align}
The domains of $u$ and $v$ are now defined by $uv<0$.
However, near $uv=0$, the metric reduces to Minkowski in the double-null coordinates;
\begin{align}
ds^2\simeq &-2dudv+\frac12 (v-u)^2d\Omega_2^2,
\end{align}
and hence analytic extension of spacetime into the region with $uv>0$ is possible.
The metric in the region with $uv>0$ is given by replacing $\sin^2(H\sqrt{-2uv})$ in the metric (\ref{EdS4}) by $-\sinh^2(H\sqrt{2uv})$.
%\begin{align}
%ds^2=&-\frac{(vdu+udv)^2}{2uv}+\frac{\sinh^2(H\sqrt{2uv})}{H^2}\frac{(vdu-udv)^2}{4u^2v^2} \nonumber \\
%&+\frac{\sinh^2(H\sqrt{2uv})}{H^2}\frac{(v-u)^2}{4uv}d\Omega_2^2.\label{EdS4} 
%\end{align}
Finally, by the transformations
\begin{eqnarray}
u=\frac{1}{\sqrt{2}}Te^{-{\bar \chi}}, ~~~v=\frac{1}{\sqrt{2}}Te^{{\bar \chi}}, \label{trans-2a}
\end{eqnarray}
of which inverse is
\begin{eqnarray}
T=\sqrt{2uv},~~~{\bar \chi}={\rm arctanh}\biggl(\frac{v-u}{v+u}\biggl), \label{trans-2b}
\end{eqnarray}
the metric (\ref{EdS4}) with $uv>0$ is transformed into 
\begin{align}
ds^2=-dT^2+a(T)^2(d{\bar \chi}^2+\sinh^2{\bar \chi}d\Omega_2^2),\label{EdS5} 
\end{align}
where $a(T)=-(1/H)\sinh HT$.
This is the de~Sitter universe in the spatially open slicing and the domain of $T$ is $T \ge 0$.
The stealth solution in the open universe is given by  
\begin{align}
\label{stealth-L1}
\phi= \left\{
\begin{array}{ll}
(-p\cosh HT+q)^{n} & (\xi\ne 1/4),\\
K e^{-w\cosh HT} & (\xi= 1/4).
\end{array} \right. 
\end{align}
The direct transformations from (\ref{EdS3}) to (\ref{EdS5}) are ${\bar \tau}=iT$ and ${\bar t}=i\pi/2+{\bar \chi}$.

The spacetime (\ref{EdS3}) is also analytically extended beyond another null surface ${\bar \tau}=\pi/H$ to the open universe in the same way.
By the transformation ${\hat \tau}:={\bar \tau}-\pi/H$, the previous argument can be used for ${\hat \tau}=0$ (namely for ${\bar \tau}=\pi/H$).
The extended spacetime is again the de~Sitter universe in the open slicing, but causally disconnected from the previous one.
In this open universe, different from (\ref{stealth-L1}), the stealth scalar field is given by 
\begin{align}
\label{stealth-L2}
\phi= \left\{
\begin{array}{ll}
(p\cosh H{\hat T}+q)^{n} & (\xi\ne 1/4),\\
K e^{w\cosh H{\hat T}} & (\xi= 1/4),
\end{array} \right. 
\end{align}
where we used the time coordinate ${\hat T}(\ge 0)$ in order to distinguish this de~Sitter region from the other one. 

As seen in Eqs.~(\ref{stealth-L1})  and (\ref{stealth-L2}), in the case of $\xi \ne 1/4$, inside the bracket in the expression of $\phi$ becomes negative at least in one of the two de~Sitter regions independent of the sign of $p$ and hence $n \in \mathbb{N}$ is required.
In addition, $\phi$ must not be negative if the power exponent in the potential is non-integer.
This requires $n=1$ or $n=2k~(k\in \mathbb{N})$, which also satisfies the regularity conditions considered in Sec.~\ref{sec:instanton}. 
We will consider $n=2k+2~(k\in \mathbb{N})$ for $q\ne |p|$ and $n=2k~(k\in \mathbb{N})$ for $q=|p|$ because the result in Sec.~\ref{sec:dS} shows that the theory admits a de~Sitter solution with the same value of $\Lambda$ and a constant scalar field in these cases. 

For $\xi=1/4$, in contrast, no additional condition is required.
As a consequence, the parameter space for creation of the open universe totally contains the one for creation of the closed universe.
In the next subsection, we will discuss which universe is preferred at
 the moment of creation.

\subsection{Creation probability with a stealth scalar field}
We have seen that, for certain values of parameters, the theory with the potential (\ref{potential-3}) admits de~Sitter solutions with or without a stealth scalar field.

In order to discuss the creation probability of the de Sitter universe,
 we evaluate the Euclidean action 
by use of the semi-classical instanton solution ($g_{\mu\nu}^{\rm (I)},
\phi_{\rm (I)}$), 
which has been discussed in the previous section:
\begin{align}
\label{action}
S_{\rm E}=&S_{\rm g(E)}+S_{\phi(E)},
\end{align}
where 
\begin{align}
S_{\rm g(E)}=&-\frac{1}{2}\int d^4x\sqrt{g^{\rm (I)}}\biggl(R(g_{\mu\nu}^{\rm (I)})
-2\Lambda\biggl) 
%\nonumber \\
%&~~~~~~~+{\rm (boundary~term)}
\,,\\
S_{\phi(E)}=&\int \D^4x \sqrt{g^{\rm (I)}}\biggl[\frac12\left(\frac{d\phi_{\rm (I)}}{d\tau}\right)^2 \nonumber \\
&~~~~~~~+\frac12\xi \phi_{\rm (I)}^2 R(g_{\mu\nu}^{\rm (I)})+V(\phi_{\rm (I)})  \biggl]
\,,
\label{action-con}
\end{align}
where $d^4 x=d\tau d^3\vect{x}$.
In the Hartle-Hawking proposal, the tunneling probability is proportional to 
$\exp(-S_{\rm E})$~\cite{footnote2}.
The probability for the solution with $\phi\equiv 0$ is simply given by 
$\exp(-S_{\rm g(E)})(=\exp(24\pi^2/\Lambda))$.
Therefore, the universe with the stealth scalar field is preferred if 
$S_{\phi(E)}(\phi_{\rm stealth})<0$.

First we compute the value of the Euclidean action for $\xi\ne 1/4$ using the coordinates (\ref{EdS2}).
The following result is valid both for the closed universe and the open universe creations.
\begin{widetext}
We compute
\begin{align}
S_{\phi(E)}=&2\pi^2\int_{0}^{\pi/H} d{\bar \tau} b^3\biggl[\frac12\left(\frac{d\phi_{\rm stealth}}{d\tau}\right)^2+2\xi \Lambda \phi_{\rm stealth}^2 +\Lambda\phi_{\rm stealth}^2\biggl(A\phi_{\rm stealth}^{(1-4\xi)/\xi}+B\phi_{\rm stealth}^{(1-4\xi)/2\xi}+C\biggl)  \biggl] \nonumber \\
=&\frac{6n\pi^2[(q+p)^{2n+1}-(q-p)^{2n+1}]}{(2n+1)(2n-1)p\Lambda }=:S_{\phi(E)}^{[\xi\ne 1/4]}.\label{intS1}
\end{align}
We are interested in the case where $n$ is a positive integer and then $S_{\phi(E)}^{[\xi\ne 1/4]}$ can be written as
\begin{align}
S_{\phi(E)}^{[\xi\ne 1/4]}=&\frac{6n\pi^2}{(2n+1)(2n-1)}\frac{p^{2n}}{\Lambda}\biggl[\biggl(1+\frac{q}{p}\biggl)^{2n+1}+\biggl(1-\frac{q}{p}\biggl)^{2n+1}\biggl]  \\
=&\frac{3^{2n}\times  6\pi^2(2n-1)^{2n-1}}{2^{2n}(2n+1)(n+1)^{2n}n^{4n-1}}\frac{B^{2n}}{\Lambda}\biggl(\frac{1+\eta}{\eta}\biggl)^n \biggl[\biggl(1\pm \sqrt{\frac{\eta}{1+\eta}}\biggl)^{2n+1}+\biggl(1\mp\sqrt{\frac{\eta}{1+\eta}}\biggl)^{2n+1}\biggl],\label{intS1-2}
%=&\frac{3^{n}\times  6\pi^2}{2^{2n}(2n+1)(2n-1)(n+1)^{2n}n^{4n-1}}\frac{[8n^2(n+1)^2A+3(2n-1)^2B^2]^n}{\Lambda} \nonumber \\
%&\times \biggl[\biggl(1\pm \sqrt{\frac{3(2n-1)^2B^2}{8n^2(n+1)^2A+3(2n-1)^2B^2}}\biggl)^{2n+1}+\biggl(1\mp\sqrt{\frac{3(2n-1)^2B^2}{8n^2(n+1)^2A+3(2n-1)^2B^2}}\biggl)^{2n+1}\biggl],
\end{align}
\end{widetext}
where $\eta$ is defined by 
\begin{align}
\eta:=&\frac{3(2n-1)^2}{8n^2(n+1)^2}\frac{B^2}{A} \label{defeta}
\end{align}
such that
\begin{align}
\frac{q^2}{p^2}=&\frac{\eta}{1+\eta}.
\end{align}
$\eta<-1$ or $\eta>0$ are required and the upper (lower) signs in front of the square root in Eq.~(\ref{intS1-2}) corresponds to $q/p>(<)0$.

In the case of the closed universe creation, we have $q=0$ and $n=4k (k\in \mathbb{N})$.
In the case of the open universe creation with $\xi \ne 1/4$, we have two cases; $n=2k+2$ for $q\ne |p|$ and $n=2k$ for $q=|p|$.
Actually $S_{\phi(E)}^{[\xi\ne 1/4]}$ is positive definite.
To show this, we use
\begin{align}
\frac{S_{\phi(E)}^{[\xi\ne 1/4]}}{S_{\phi(E)}|_{q=0}}=&\frac12\biggl[\biggl(1+\frac{q}{p}\biggl)^{2n+1}+\biggl(1-\frac{q}{p}\biggl)^{2n+1}\biggl],\label{eval1}
\end{align}
where $S_{\phi(E)}|_{q=0}$ is the value of $S_{\phi(E)}^{[\xi\ne 1/4]}$ with $q=0$:  
\begin{align}
S_{\phi(E)}|_{q=0}:=\frac{3^n\times 12\pi^2 n }{(2n-1)(2n+1)(n^2/2)^{n}}\frac{A^{n}}{\Lambda}(>0). \label{Sq=0}
\end{align}
Since the right-hand side of Eq.~(\ref{eval1}) is an even function and greater than $1$ for $q\ne 0$, $S_{\phi(E)}^{[\xi\ne 1/4]}\ge S_{\phi(E)}|_{q=0}>0$ is satisfied.
Namely, $S_{\phi(E)}^{[\xi\ne 1/4]}$ is positive and bounded from below by $S_{\phi(E)}|_{q=0}$.
Therefore, the creation of the universe with the stealth scalar field is not preferred.

However, the creation rate of the de Sitter universe with a stealth scalar field can be almost the same as that without it.
In order to show how plausible such a creation is, we shall evaluate $S_{\phi(E)}^{[\xi\ne 1/4]}$.
Since we discuss creation of the universe, we expect the cosmological constant $\Lambda\simeq \mathcal{O}(1)$ and the coefficients in the potential $A, B, C \simeq \mathcal{O}(1)$ in the Planck unit.
Actually, the value of $\exp(-S_{\phi(E)}^{[\xi\ne 1/4]})$ is very sensitive to the parameters and can be close both to $0$ and $1$ even with such parameters.

Lastly we compute $S_{\rm g(E)}$ for $\xi=1/4$, which is used only for the open universe creation.
We obtain 
\begin{widetext}
\begin{align}
S_{\phi(E)}=&2\pi^2\int d{\bar \tau} b^3\biggl[\frac12\left(\frac{d\phi_{\rm stealth}}{d\tau}\right)^2+2\Lambda \xi \phi_{\rm stealth}^2 +\Lambda\phi_{\rm stealth}^2\biggl({\bar A}+{\bar B}\ln {\phi}_{\rm stealth}-\frac{1}{6}(\ln{\phi}_{\rm stealth})^2\biggl) \biggl] \nonumber \\
=&\frac{3\pi^2e^{6{\bar A}-w^2}}{\Lambda} \biggl[\frac{2\{w^2-(6{\bar A}-1)\}^2+1}{2w^2}\biggl(\cosh(2w) -\frac{\sinh(2w) }{2w}\biggl)+\frac{\sinh(2w)}{w}\biggl]=:S_{\phi(E)}^{[\xi= 1/4]}.\label{intS2}
\end{align}
\end{widetext}
It is easy to show $S_{\phi(E)}^{[\xi= 1/4]}>0$ and hence the creation of the de~Sitter universe with a stealth scalar field is not preferred also in this case.
However, again the creation rate of the open universe with a stealth scalar field sharply depends on the parameters and $\exp(-S_{\phi(E)}|_{\xi= 1/4})$ can be close both to $0$ and $1$ with ${\bar A}, {\bar B}\simeq \mathcal{O}(1)$.

%======================================%
%<<<<<<<<<<<< SECTION IV >>>>>>>>>>>>>>%
%======================================%
\section{Summary}
\label{sec:summary}
In this paper, we have studied the contribution of a stealth scalar field 
to the creation probability of the de~Sitter universe.
With a certain form of the potential, we have constructed an exact stealth solution in the Hawking-Moss de~Sitter
 instanton space.
An analytic continuation of the instanton with a stealth scalar field to the de~Sitter universe is allowed for a discrete value of the coupling constant.
While such a continuation is possible to the de~Sitter universe both in the spatially closed and open slicing for $\xi \ne 1/4$, it is allowed only to the spatially open universe for $\xi=1/4$.

Adopting the Hartle-Hawking proposal, we have shown that the creation probability with a stealth scalar field is always less than that with a trivial scalar field.
However, the creation of the de~Sitter universe with a stealth scalar field is quite sensitive to the parameters and can be almost the same as that of the universe in vacuum.
Actually, the creation probability with a stealth scalar field can be larger if we adopt the Linde-Vilenkin proposal.
Nevertheless, the effect of the stealth field is significant both in the frameworks of Hartle-Hawking and Linde-Vilenkin proposals if $|S_{\phi(E)}/S_{\rm g(E)}|\ll1$ is satisfied.

In the present paper, we have considered only the Hawking-Moss 
de~Sitter instanton.
However, one may expect that there exist other types of instantons.
Examples are the Halliwell-Laflamme instanton for a conformally coupled scalar field ($\xi=1/6$) without potential~\cite{hl1989} and the Coleman-De~Luccia instanton or Hawking-Turok instanton~\cite{cdl1980,ht1998} for a minimally coupled scalar field.
If it is the case, we have to take into account such instantons as well,
although we suspect that the highly symmetric de~Sitter instanton is most preferable.
We shall leave it as a future problem.

\subsection*{Acknowledgements}
HM thanks Julio Oliva for pointing out the reference~\cite{dAFF1976}.
HM would like to thank DAMTP, University of Cambridge, for its hospitality while part
of this work was carried out.
KM would like to acknowledge CECs for hospitality at his visit, where 
this work was started.
This work has been funded by the Fondecyt grants 1100328, 1100755 (HM) and 
by the Conicyt grant "Southern Theoretical Physics Laboratory" ACT-91. 
This work was also partly supported by the JSPS Grant-in-Aid for Scientific 
Research (A) 22244030 and (C) 22540291.
The Centro de Estudios Cient\'{\i}ficos (CECs) is funded by the Chilean 
Government through the Centers of Excellence Base Financing Program of 
Conicyt.

%\appendix

%\section{Halliwell-Laflamme instanton}
%In the case of a conformally coupled scalar field $\xi=1/6$ without potential, there is the Halliwell-Laflamme instanton, given by 
%\begin{align}
%a(\tau)=&\sqrt{\frac{3}{2\Lambda}\biggl(1-\sqrt{1+\frac23E\Lambda}\cos\sqrt{\frac{4\Lambda}{3}}\tau\biggl)},\label{hl}
%\end{align} 
%where $E$ is the constant in the master equation for $X(\tau):=a(\tau)^2$:
%\begin{align}
%E=&\frac12 {X'}^2+U(X),\\
%U(X):=&-2X+\frac23\Lambda X^2.
%\end{align} 
%The equation for $\phi$ is
%\begin{align}
{%(a\phi)'}^2=\frac{a^2\phi^2+3E}{a^2},
%\end{align}
%which is difficult to solve to obtain $\phi$ in a closed form.
%The metric (\ref{EdS}) with the scale factor (\ref{hl}) is non-singular and periodic for $-3/(2\Lambda)<E<0$, corresponding to the Halliwell-Laflamme instanton.
%This Halliwell-Laflamme instanton can be analytically continued to the Lorentzian spacetime at $\tau=0$, providing a de~Sitter-like universe.

\end{document}